\documentclass[11pt]{article}
\usepackage{moriond,epsfig,rotating}

\def\be{\begin{equation}}
\def\ee{\end{equation}}
\def\bea{\begin{eqnarray}}
\def\eea{\end{eqnarray}}

\newcommand{\gev} {{\,\rm GeV\/}}
\newcommand{\PTM} {P_{T,\rm miss}}
\newcommand{\coll}{Coll.}
\newcommand{\etal}{et al.}
\newcommand{\til}{~}
\newcommand{\Lumi}{{\cal L}}
\newcommand{\pb}{\,\rm {pb}}
\newcommand{\pbi}{\,\rm {pb}^{-1}}

\begin{document}
\vspace*{4cm}
\title{Measurement of high-\boldmath{$Q^2$} deep inelastic scattering cross sections 
with longitudinally polarised positron beams at HERA}

\author{ Julian Rautenberg\\ On behalf of the H1
 and ZEUS collaborations}

\address{ Physikalisches Institut, Universit\"at Bonn, Nussallee 12,
53115 Bonn, Germany\\
supported by the German Federal Ministry for Education and Research (BMBF)}

\maketitle\abstracts{
The first measurements of the cross sections for neutral and charged current 
deep inelastic scattering in $e^{+}p$ collisions with longitudinally polarised 
positron beams are presented. The total cross section for $e^+p$ charged 
current deep inelastic scattering is presented at positive and negative 
values of positron beam longitudinal polarisation 
for an integrated luminosity of $37.0 \pbi$ H1 data and  $30.5 \pbi$ ZEUS data
collected in 2003 and 2004 at a centre-of-mass energy of 319 \gev. 
In addition, the ZEUS collaboration measured the single differential
cross sections for charged and neutral current deep inelastic
scattering in the kinematic region $Q^{2}>200 \gev^{2}$.
The measured cross sections are compared
 with the predictions of the Standard Model.
The H1 collaboration extrapolate the cross section to a fully left handed positron beam
and find it to be consistent with the Standard Model expectation.
}


\section{Introduction}
\label{s:intro}

Deep inelastic scattering (DIS) of leptons off nucleons has proved to
be a key tool in the understanding of the structure of the proton and
the form of the Standard Model (SM).  
The HERA $ep$ collider has made possible 
the exploration of DIS at high values of negative four-momentum-transfer squared, $Q^{2}$. 
Using data taken in the years 1994-2000 the H1 and ZEUS collaborations have reported
measurements of the cross sections for charged current (CC)
and neutral current (NC)
DIS~\cite{Adloff:2003uh,Chekanov:2003vw}.
These measurements extend the kinematic region covered by fixed-target
experiments~\cite{zfp:c25:29} to
higher $Q^{2}$ and allow the 
HERA experiments to probe the electroweak sector of the SM. 

This paper presents the measurements of the cross sections for
$e^+ p$ CC and NC DIS 
with longitudinally polarised positron beams. 
The measurements are based on the integrated luminosities 
collected at the luminosity weighted mean polarisations given in Table~\ref{tab:lumpol}
with the ZEUS and H1 detectors in 2003 and 2004. During this time HERA
collided protons of energy 
$920 \gev$ with positrons of energy $27.6 \gev$, yielding collisions
at a centre-of-mass energy of $319 \gev$. 
The measured cross sections are compared to the SM predictions.

\begin{table}[btp]
  \centering
\begin{tabular}{|c|c|c|}
\hline
&$\Lumi/\pbi$&$P$\\\hline
 H1   & 15.3  & +33.0\%\\
     & 21.7  & -40.2\%\\\hline
 ZEUS   & 14.1 & +31.8\%\\
     & 16.4 & -40.2\%\\\hline
\end{tabular}
  \caption{Integrated luminosities 
    and the luminosity weighted mean longlitudinal polarisation of the lepton beam 
    for the H1 and ZEUS data samples.}
  \label{tab:lumpol}
\end{table}


\section{\bf Polarised lepton beams}
\label{s:pol}

\begin{figure}[b]
      \begin{center}
        \psfig{figure=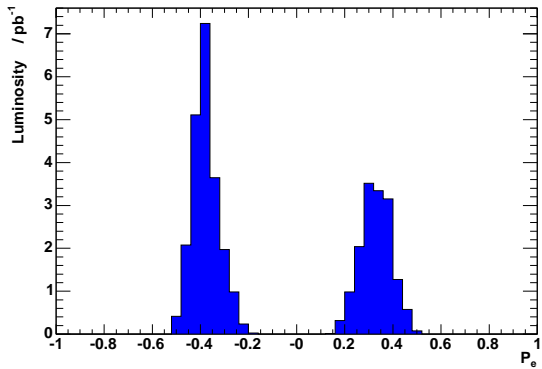,width=0.5\textwidth}
      \end{center}
      \caption{Distribution of luminosity versus the polarisation of the lepton beam, $P_e$, for the H1 data sample.}
      \label{fig:H1pol}
    \end{figure}

At HERA transverse polarisation of the lepton beam arises through synchrotron radiation 
via the Sokolov-Ternov effect~\cite{Sokolov:1963zn}.
As a part of the HERA upgrade in the year 2000 spin rotators
that rotate the polarisation into the longitudinal direction 
were installed in the lepton beamline round the collision regions where the H1 and ZEUS 
detectors are located. The polarisation is continuously measured using two independent 
polarimeters. The TPOL~\cite{Barber:1992fc} is situated at a position of transverse lepton beam polarisation,
and the LPOL~\cite{Beckmann:2000cg} at one of longitudinal polarisation. 
The luminosity weighted polarisation distribution for the data 
of the presented H1 measurement is shown in Fig.~\ref{fig:H1pol}.

%
%
\section{\bf Kinematic variables and cross sections}
\label{s:Kincross}

Inclusive deep inelastic lepton-proton scattering can be described in terms of the kinematic 
variables $x$, $y$ and $Q^2$.
The variable $Q^2$ is defined to be $Q^2 = -q^2 = -(k-k')^2$ where $k$ and $k'$ 
are the four-momenta of the incoming and scattered lepton, respectively. Bjorken $x$ is defined
by $x=Q^2/2P \cdot q$ where $P$ is the four-momentum of the incoming proton. The 
variable $y$ is defined by $Q^2=sxy$, where $s=4E_e E_p$ is the square 
of the lepton-proton centre-of-mass energy (neglecting the masses of the incoming particles).

The electroweak Born level cross section for the CC reaction, $e^{+}p\rightarrow \bar{\nu_{e}}X$, with a longitudinally polarised positron beam (defined in Eqn.~(\ref{eqn:pol})), can be expressed as~\cite{proc:hera:1995:163}

\begin{equation}
\frac{d^2 \sigma ^{CC} (e^+ p)}{dxdQ^2}=(1+\mathcal{P})\frac{G_{F}^{2}}{4\pi x}\bigg(\frac{M_{W}^{2}}{M_{W}^{2}+Q^{2}}\bigg) ^{2} \cdot \bigg[ Y_{+}F_{2}^{CC}(x,Q^{2})-Y_{-}xF_{3}^{CC}(x,Q^{2}) -y^{2}F_{L}^{CC}(x,Q^{2})\bigg], \nonumber
\end{equation}

where $G_{F}$ is the Fermi constant, $M_{W}$ is the mass of the $W$ boson and $Y_{\pm}=1\pm(1-y)^{2}$. The structure functions $F_{2}^{CC}$ and $xF_{3}^{CC}$ contain sums and differences of the quark and anti-quark parton density functions (PDFs) and $F_{L}^{CC}$ is the longitudinal structure function. The longitudinal polarisation of the positron beam is defined as

\begin{equation}
\mathcal{P}=\frac{N_{R}-N_{L}}{N_{R}+N_{L}},
\label{eqn:pol}
\end{equation}
  
where $N_{R}$ and $N_{L}$ are the numbers of right and left-handed positrons in the beam. Similarly the cross section for the NC reaction, $e^{+}p\rightarrow e^{+}X$, can be expressed as

\begin{equation}
\frac{d^2 \sigma ^{NC} (e^+ p)}{dxdQ^2}=\frac{2\pi \alpha ^2}{xQ^4} [H_{0}^{+}+\mathcal{P}H_{\mathcal{P}}^{+}], \nonumber
\end{equation}

where $\alpha$ is the QED coupling constant and $H_{0}^{+}$ and $H_{\mathcal{P}}^{+}$ contain the unpolarised and polarised structure functions, respectively.

Charged current events are characterised by a large missing transverse momentum, $\PTM$, 
which is calculated by ZEUS as 

\begin{equation}
\PTM^2  =  P_x^2 + P_y^2 = 
  \left( \sum\limits_{i} E_i \sin \theta_i \cos \phi_i \right)^2
+ \left( \sum\limits_{i} E_i \sin \theta_i \sin \phi_i \right)^2, \nonumber
  \label{eq:pt}
\end{equation}

where the sum runs over all calorimeter energy deposits $E_i$, and
$\theta_i$ and $\phi_i$ are the polar and azimuthal angles
of the calorimeter cell as viewed from the interaction vertex.
The hadronic polar angle, $\gamma_h$, is defined by
$\cos\gamma_h = (\PTM^2 - \delta^2)/(\PTM^2 + \delta^2)$,
where
$\delta = \sum ( E_i - E_i \cos \theta_{i} ) 
= \sum (E-P_z)_{i}$. 
H1 uses an analog definition summing over all final state particles. 
In the naive Quark Parton Model,
$\gamma_h$ gives the scattering angle of the struck quark in the laboratory frame.
The total transverse energy,
$E_T$, is given by
$E_T    = \sum E_i \sin \theta_i$. 

Neutral current events are characterised by the presence of a high-energy isolated scattered positron in the detector.
It follows from longitudinal momentum conservation that for well measured NC events $\delta$ peaks at twice the positron beam 
energy i.e. $55 \gev$. 

The kinematic variables for charged current events were reconstructed from the measured $\PTM$ and $\delta$
using the Jacquet-Blondel method \cite{proc:epfacility:1979:391}.
For the neutral current events 
the ZEUS measurement uses the  double-angle method~\cite{proc:hera:1991:23} 
to estimate the kinematic variables from the polar angles of the scattered positron 
$\theta_e$
and the hadronic final state $\gamma_h$
while H1 uses the $e\Sigma$ method~\cite{nim:a361:197} to estimate the kinematic quantities 
from the scattered positrons energy, its angle and $\delta$.


\section{Monte Carlo simulation}
\label{s:Simu}

Monte Carlo simulation (MC) was used to determine the efficiency for 
selecting events, the accuracy of kinematic 
reconstruction, to estimate the background rate and to deduce
cross sections for the full kinematic region
from the data.
A sufficient number of
events were generated to ensure that uncertainties from MC 
statistics were small.
The MC samples were normalised to the total
integrated luminosity of the data.

Neutral and charged current DIS events including radiative effects were simulated 
using the {\sc djangoh}~\cite{proc:hera:1991:1419} generator.
The hadronic final state was simulated
using the colour-dipole model of {\sc ariadne}~\cite{cpc:71:15}.
For the hadronisation the Lund string model of {\sc jetset} 7.4~\cite{cpc:39:347}
is used. 
Additional samples used are as described in~\cite{Adloff:2003uh,Chekanov:2003vw}.


\section{Event selection}
\label{s:xsec}

\begin{figure}[tb]
  \begin{minipage}[c]{0.495\textwidth}
    \begin{turn}{270}
\psfig{figure=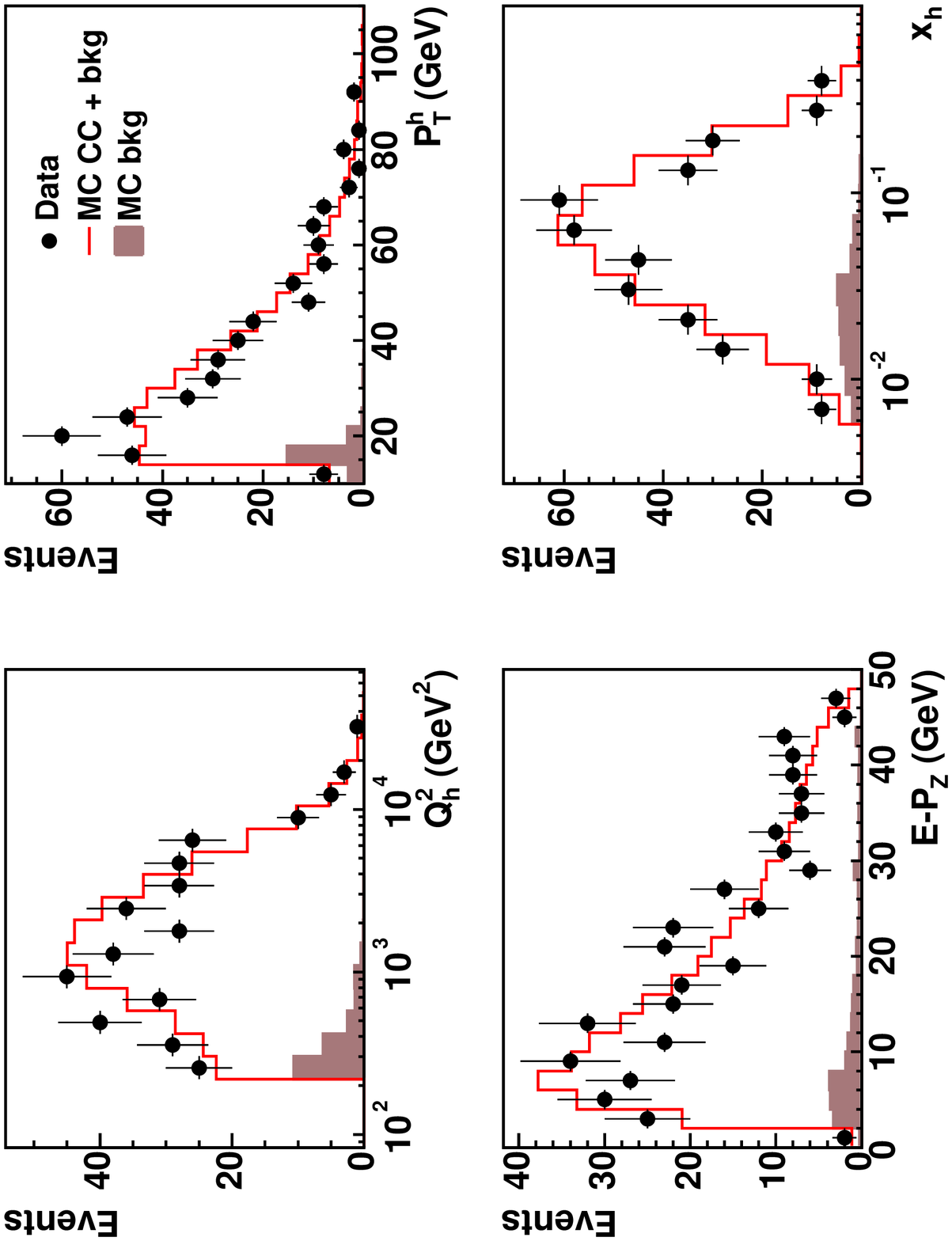,height=\textwidth}      
    \end{turn}
  \end{minipage}\hfill
  \begin{minipage}[c]{0.46\textwidth}
\psfig{figure=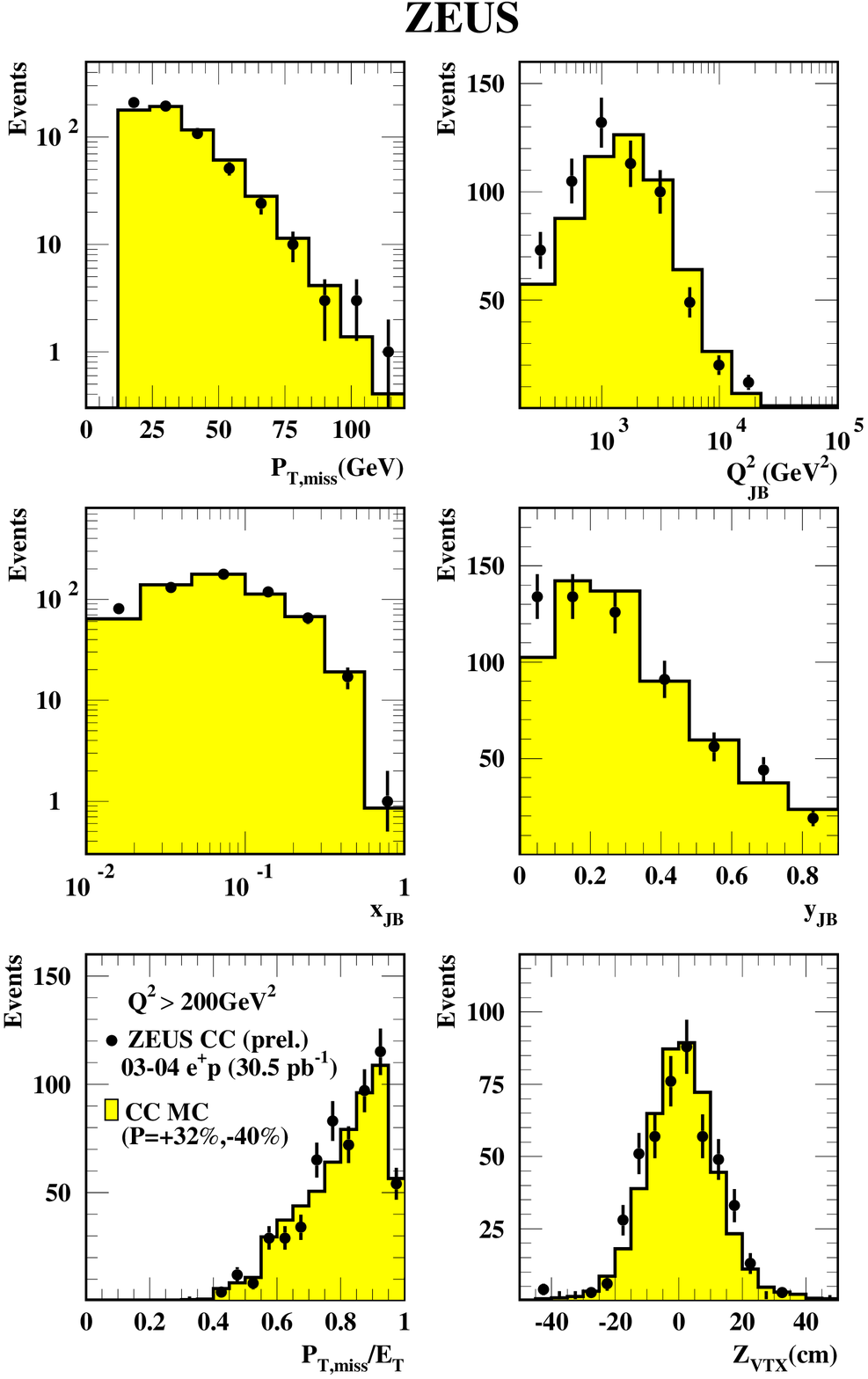,width=\textwidth}
  \end{minipage}%
\caption{Comparison of the final $e^+p$ CC data samples 
with the expectation of the Monte Carlo simulation described in the text.
On the left the distributions $Q^2$, $\PTM$, $E-P_z$ and $x$ for the right handed H1 CC data sample
are shown
and on the right $\PTM$, $Q^2$, $x$, $y$, $\PTM / E_T$ and $Z_{\rm VTX}$ 
for the ZEUS CC data sample.}
\label{fig:cc-ctrl}
\end{figure}

CC events are selected by requiring $\PTM > 12\gev$.
In order to ensure high trigger efficiency and good kinematic resolution
the analysis is restricted to the kinematic region of $Q^2>200\gev^2$
and in $y$ by $0.03<y<0.85$ for the H1 measurement and $y<0.9$ for the ZEUS measurement.
Non-$ep$ background is rejected by searching for typical beam-induced
background event topologies.
For the ZEUS measurement a total of 604 candidate events 
passed the selection criteria. 
The background contamination was estimated to 
be typically less than 1\% but was as high as 5\% in the lowest $Q^{2}$ bin of the negative polarisation sample. 
The contribution of simulated background for the H1 measurement is
visible at low $Q^2$ in Fig.~\ref{fig:cc-ctrl}.
It shows a 
comparison of data and MC distributions for the CC sample of the H1
and ZEUS measurements. The Monte Carlo gives a good 
description of the data.


\begin{figure}[tb]
  \begin{minipage}[c]{0.495\textwidth}
    \begin{turn}{270}
\psfig{figure=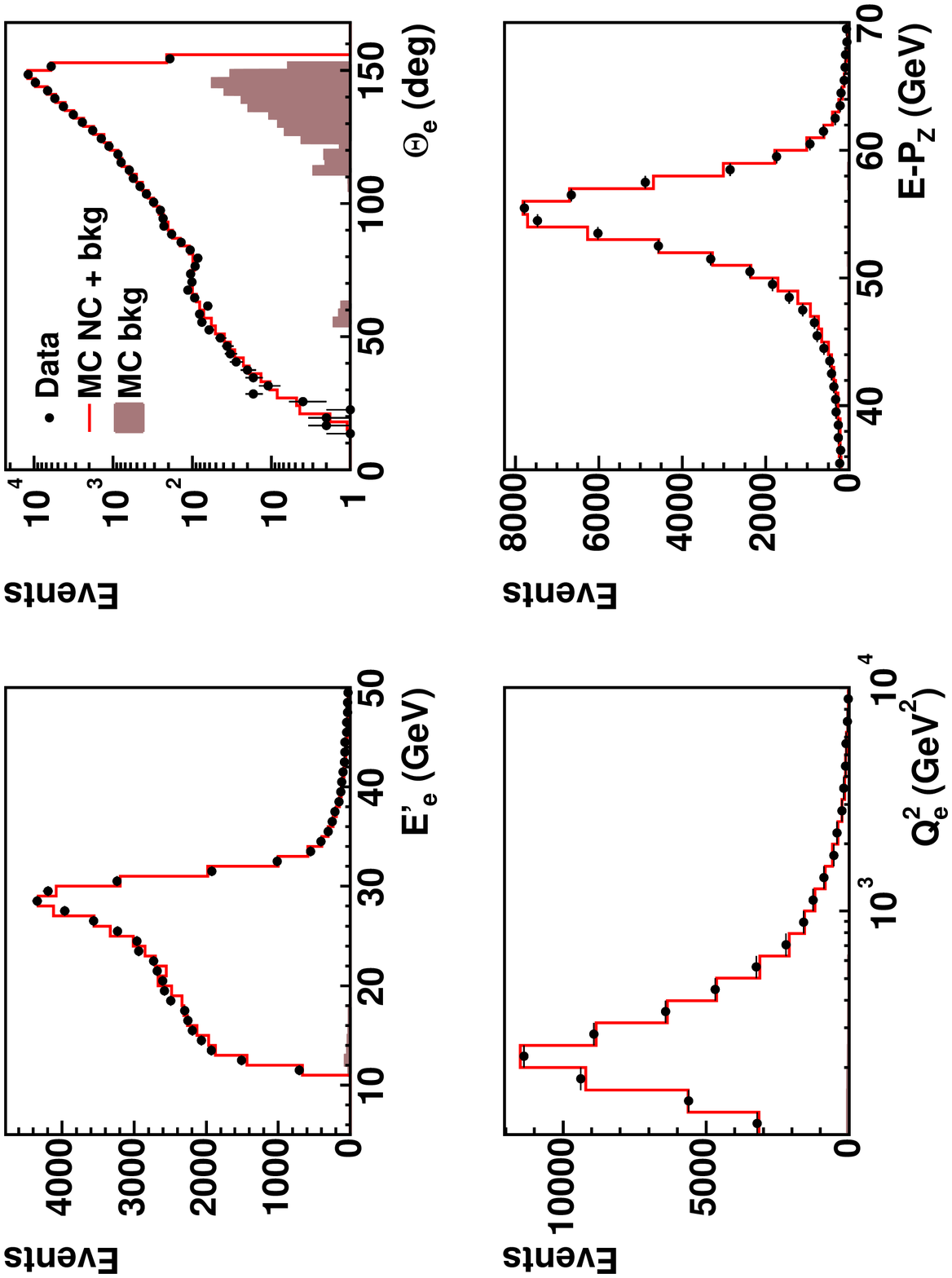,height=\textwidth}
    \end{turn}
  \end{minipage}\hfill
  \begin{minipage}[c]{0.454\textwidth}
\psfig{figure=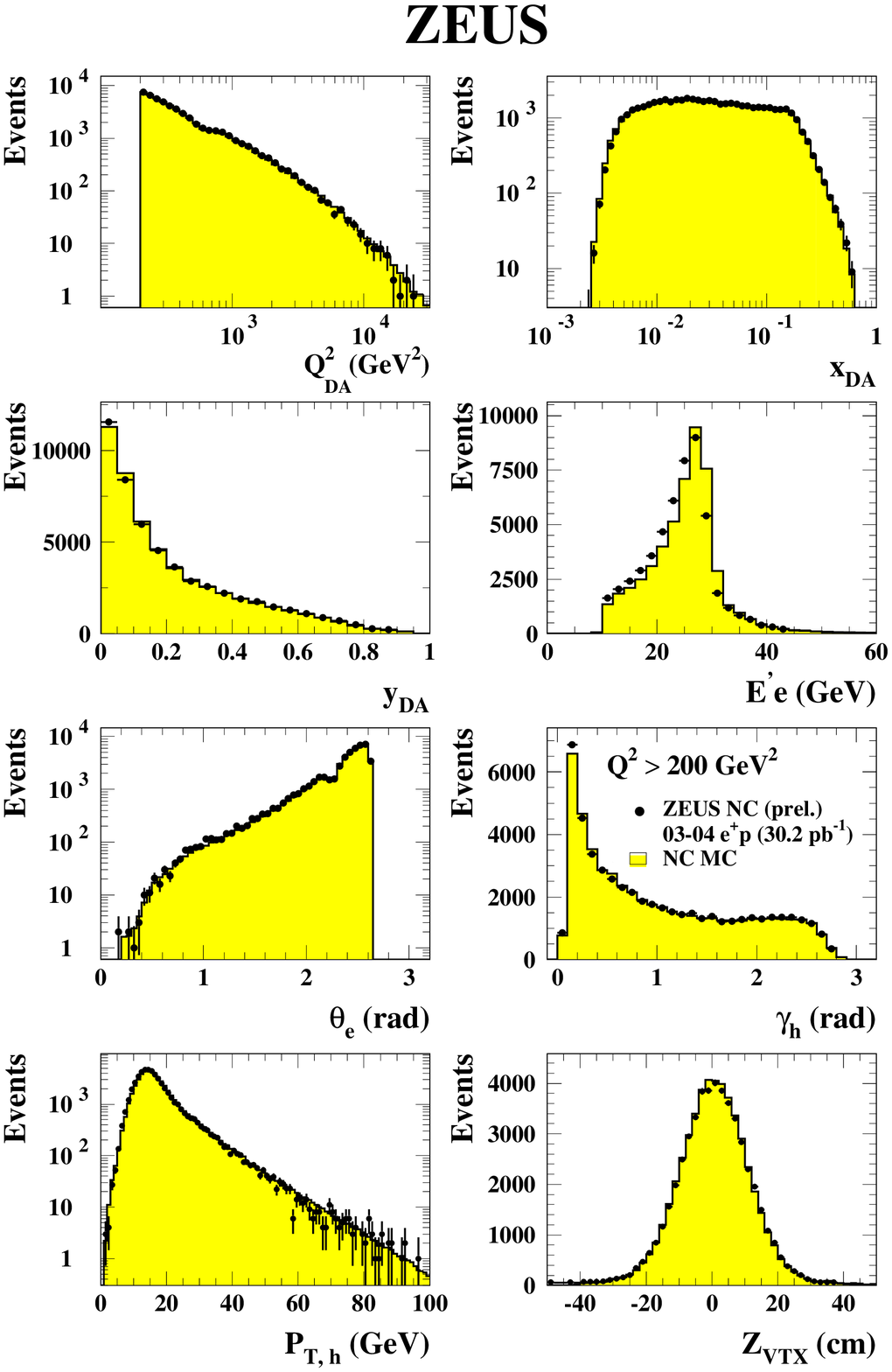,width=\textwidth}
  \end{minipage}%
\caption{Comparison of the final $e^+p$ NC data samples 
with the expectation of the Monte Carlo simulation described in the text.
On the left the distributions $E_e$, $\theta_e$, $Q^2$ and $E-P_z$ for the right handed H1
NC data sample
and on the right $Q^2$, $x$, $y$, $E_e'$, $\theta_e$, $\gamma_h$, $P_{t,h}$ and $Z_{\rm VTX}$ for the ZEUS NC data sample
are shown.
\label{fig:nc-ctrl}}
\end{figure}

NC events are selected by identifying the scattered electron. 
For this task sophisticated algorithms are used by both experiments.
The main background supression is achieved by requiring a reconstructed scattered electron energy  
$E_e'> 11 \gev$ for H1 and $E_e'>10 \gev$ for ZEUS with additional isolation criteria.
The ZEUS analysis selected a total of 52004 candidate events in the kinematic region of  $Q^2>200\gev^2$ and $y<0.95$.
The background contamination was estimated to 
be typically less than 1\%. Figure~\ref{fig:nc-ctrl} shows a 
comparison of data and MC expectation distributions for the NC sample. The Monte Carlo gives a generally 
good description of the data. Inaccuracies in the simulation of the scattered positron energy distribution
are considered in the corresponding systematic uncertainty. The effects of the positron fiducial-volume cut 
can be seen in the distribution of the scattered positron angle.

\section{Results}
\label{ss:Results}

The total cross section for $e^+ p$ CC DIS in the kinematic region $Q^{2}>200 \gev^{2}$ was measured 
by ZEUS at the given longitudinal
positron beam polarisations to be:
\begin{eqnarray}
\sigma_{\rm CC}(P=31.8 \pm 0.9 \%) &=&46.7 \pm 2.4 ({\rm stat.}) \pm 1.0 ({\rm syst.}) {\rm pb}, \nonumber \\
\sigma_{\rm CC}(P=-40.2 \pm 1.1 \%) &=&22.5 \pm 1.6 ({\rm stat.}) \pm 0.5 ({\rm syst.}) {\rm pb}. \nonumber
\end{eqnarray}

\begin{figure}[tbp]
  \begin{minipage}[c]{0.47\textwidth}
\psfig{figure=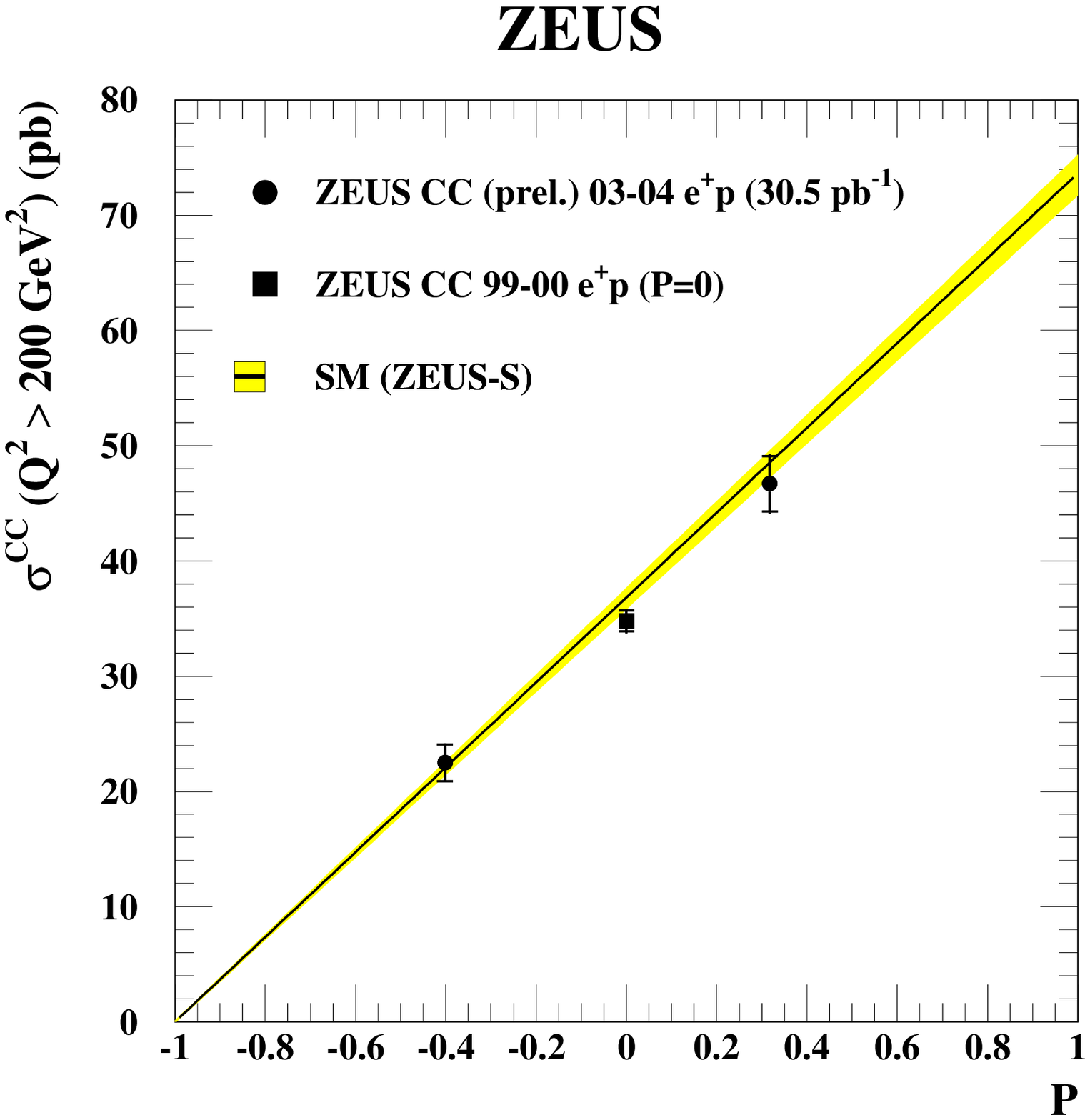,width=\textwidth}
  \end{minipage}%
  \begin{minipage}[c]{0.52\textwidth}
\psfig{figure=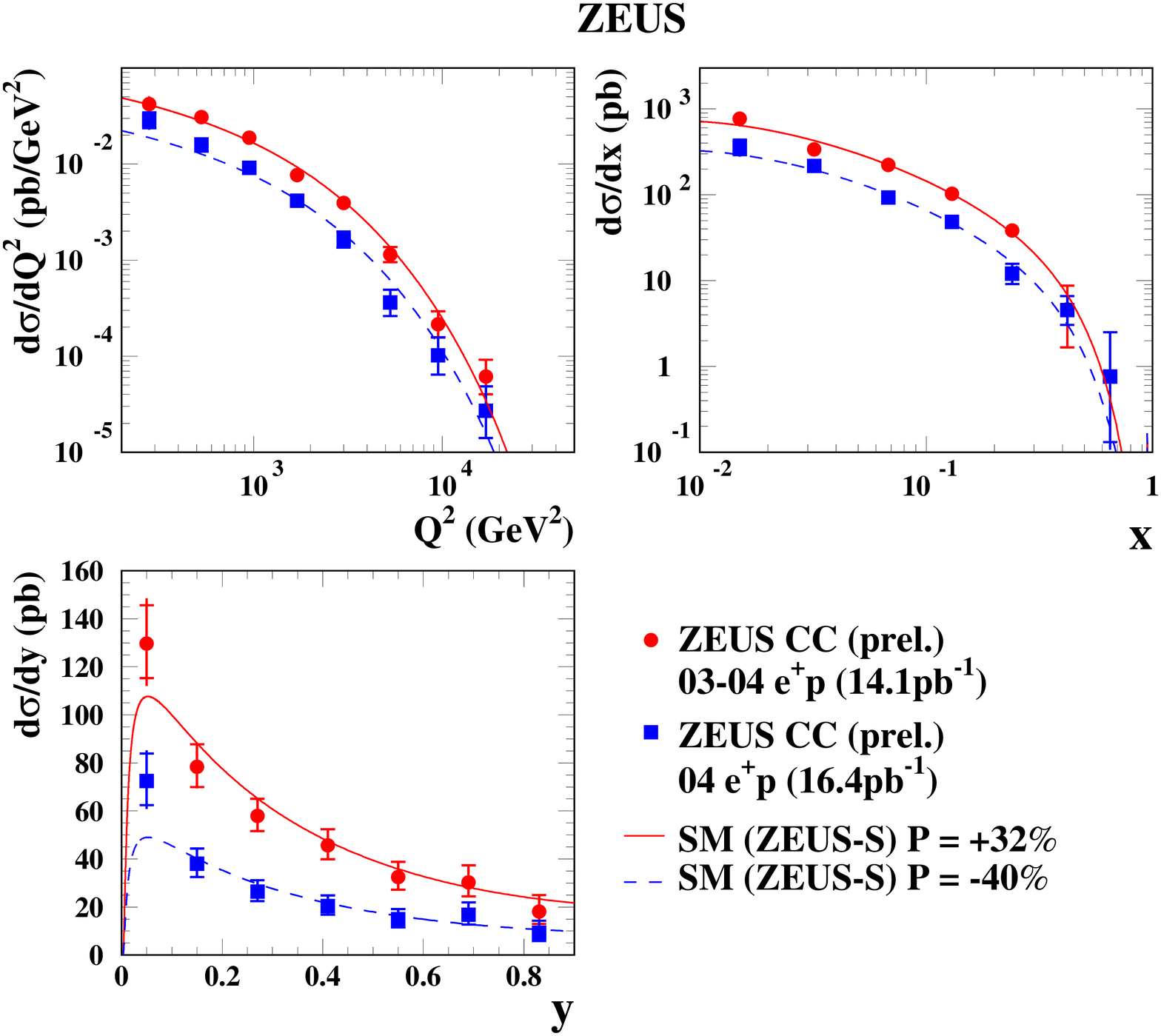,width=\textwidth}
  \end{minipage}\hfill
\caption{The total cross sections measured by ZEUS are shown on the left 
against the longitudinal polarisation of the lepton beam.
On the right 
the $e^+p$ DIS CC cross-sections $d\sigma / dQ^2$, $d\sigma / dx$ and $d\sigma / d y$
are shown.
The circles (squares) represent the data for positive (negative) polarisation 
measurements. 
The curves show the SM prediction evaluated using the ZEUS-S PDFs.
\label{fig:cc-zeus}}
\end{figure}
Note that the ZEUS measurement does not include 
the uncertainty in the luminosity of $\pm5\%$ in the systematic uncertainty.
The total cross section is shown as a function of 
the longitudinal polarisation of the positron beam in Fig.~\ref{fig:cc-zeus} including the unpolarised ZEUS 
measurement from the 1999-2000 data~\cite{Chekanov:2003vw}. 
The data are compared to the SM prediction evaluated using the ZEUS-S PDFs~\cite{pr:d67:012007}. 
The SM prediction describes the data well. The cross section points at polarisations of 31.8\% and -40.2\% respectively are 
3.4 standard deviations above and 6.1 below the unpolarised measurement, respectively.

The single-differential cross-sections, $d\sigma/dQ^{2}$, $d\sigma/dx$ and $d\sigma/dy$ for charged current DIS 
are also shown in Fig.~\ref{fig:cc-zeus}. A clear difference
is observed between the measurements for positive and negative longitudinal polarisation, 
which is well described over the whole kinematic region by the SM evaluated using the ZEUS-S PDFs.

\begin{figure}[tbp]
  \begin{minipage}[c]{0.47\textwidth}
\psfig{figure=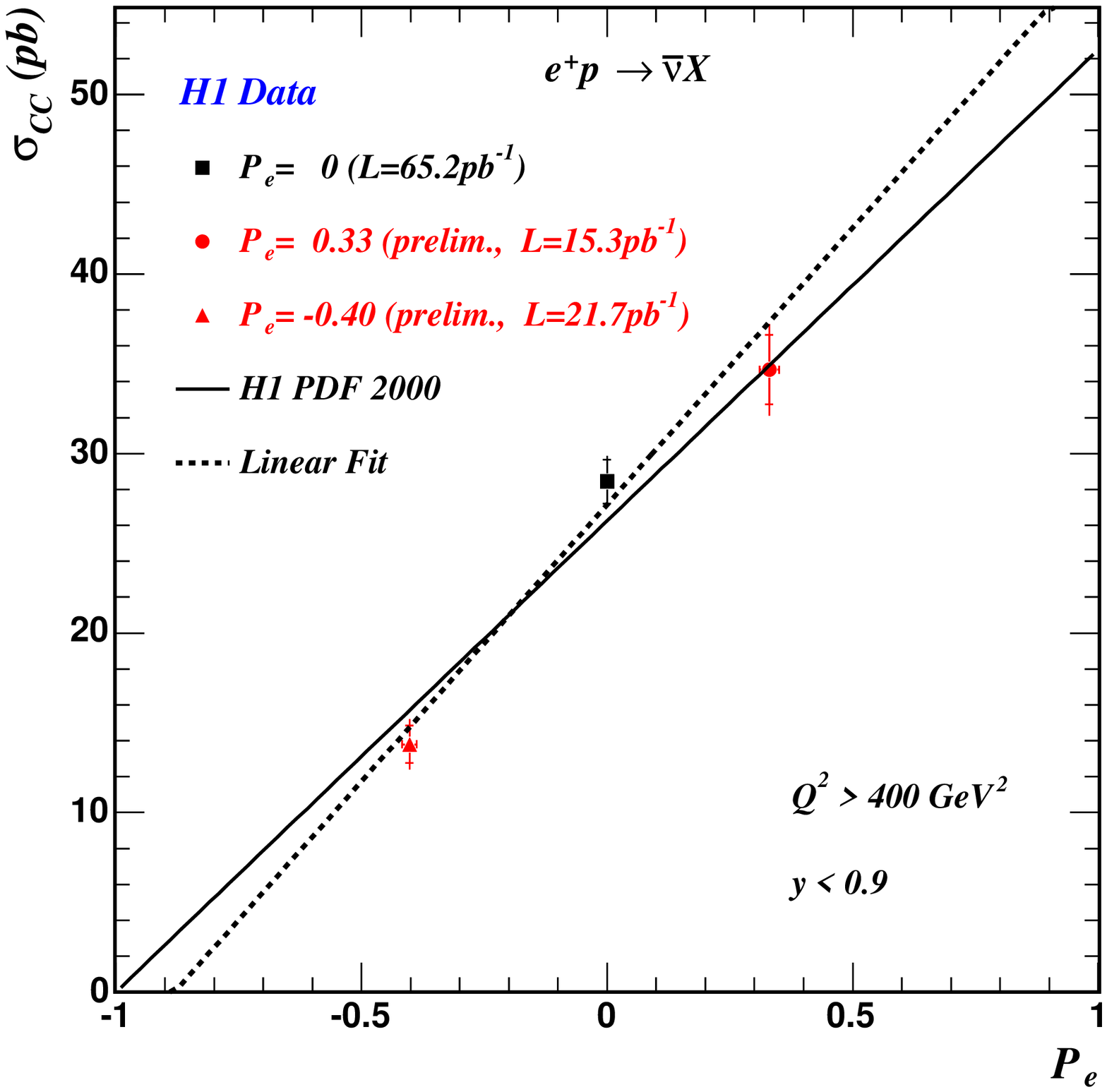,width=\textwidth}
  \end{minipage}\hfill
  \begin{minipage}[c]{0.50\textwidth}
\psfig{figure=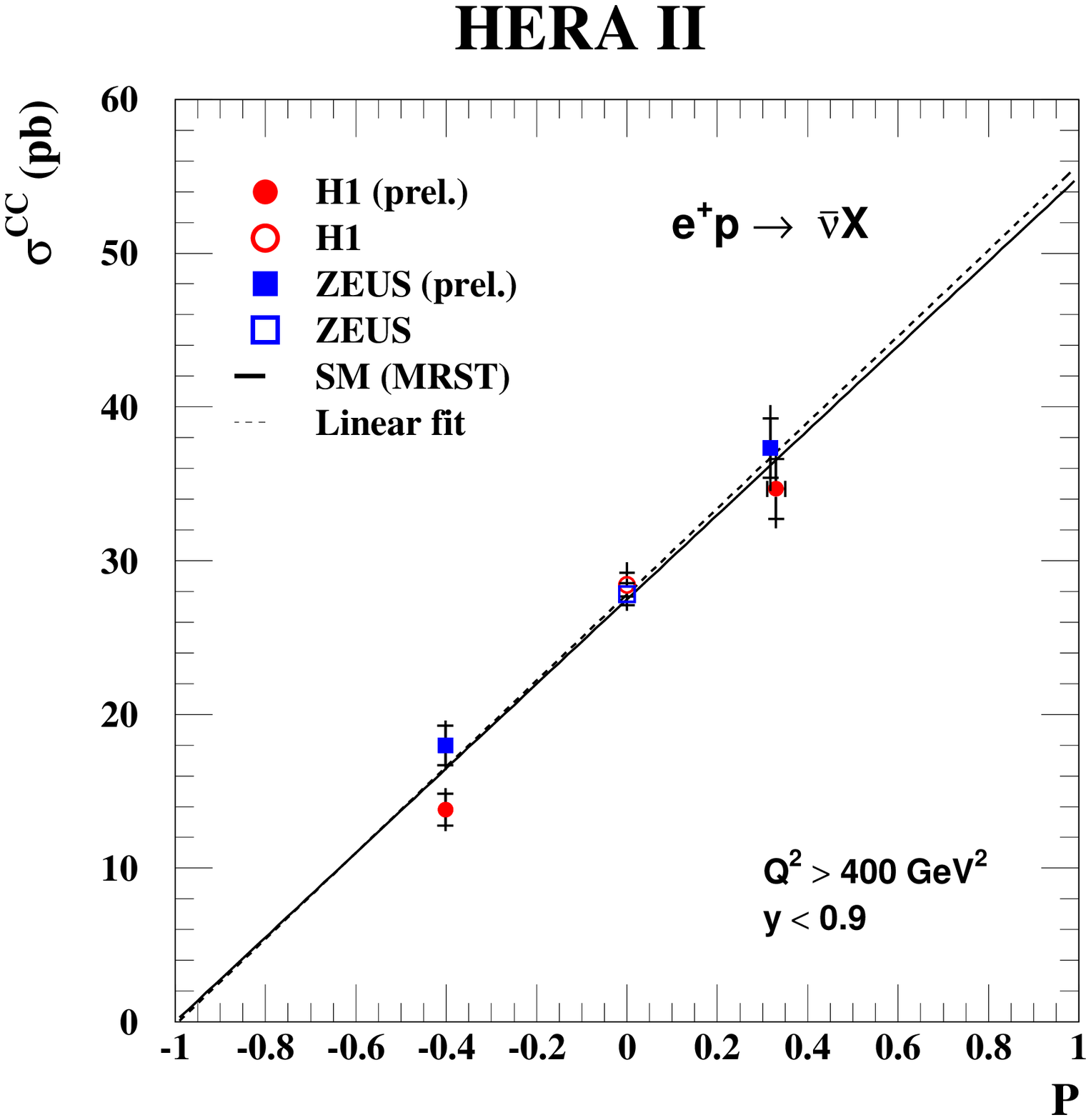,width=\textwidth}
  \end{minipage}%
\caption{On the left the total $e^+p$ DIS CC cross section measured by H1 is plotted against the 
longitudinal lepton beam polarisation. The full line shows the SM prediction evaluated using the
H1 PDF 2000 fit. The dashed line shows the result of a linear fit to the data. On the right the 
H1 measurement is compared to the ZEUS result scaled to the same kinematic region. 
\label{fig:cctot-h1}}
\end{figure}

H1 measured the integrated polarised cross section in the kinematic range 
$Q^2>400\gev^2$ and $y<0.9$ to be:
\begin{eqnarray}
  \label{eq:cctot-h1}
    \sigma_{\rm CC} (P=+33\pm 2 \%) &=& 34.7 \pm 1.9 ({\rm stat.})\pm 1.7 ({\rm syst.}) \pb, \nonumber\\
    \sigma_{\rm CC} (P=-40.2\pm 1.5 \%) &=& 13.8 \pm 1.0 ({\rm stat.})\pm 1.0 ({\rm syst.}) \pb. \nonumber
\end{eqnarray}
The total cross section is shown as a function of the longitudinal polarisation together with the 
unpolarised CC cross section measurement based on the HERA I data set~\cite{Adloff:2003uh} with an integrated luminosity of 
$65.2\pbi$ in Fig.~\ref{fig:cctot-h1}. These are compared to the SM expectation 
using the H1 PDF 2000 fit~\cite{Adloff:2003uh}. The same figure shows the result of a linear fit to the dependence of 
the cross section on the longitudinal polarisation describing reasonably the data.
Extrapolating the fit to the point $P=-1$ yields:
\begin{equation}
  \label{eq:h1wl}
      \sigma_{\rm CC} (P=-1) = -3.7 \pm 2.4 ({\rm stat.})\pm 2.7 ({\rm syst.}) \pb. \nonumber
\end{equation}
This extrapolation is consistent with the SM prediction of zero cross section.
The comparison of the H1 results to the ZEUS results scaled to the
kinematic region corresponding to the H1 measurement in
Fig.~\ref{fig:cctot-h1}
shows a good agreement between the two experiments.

\begin{figure}[!b]
  \begin{minipage}[c]{0.455\textwidth}
\psfig{figure=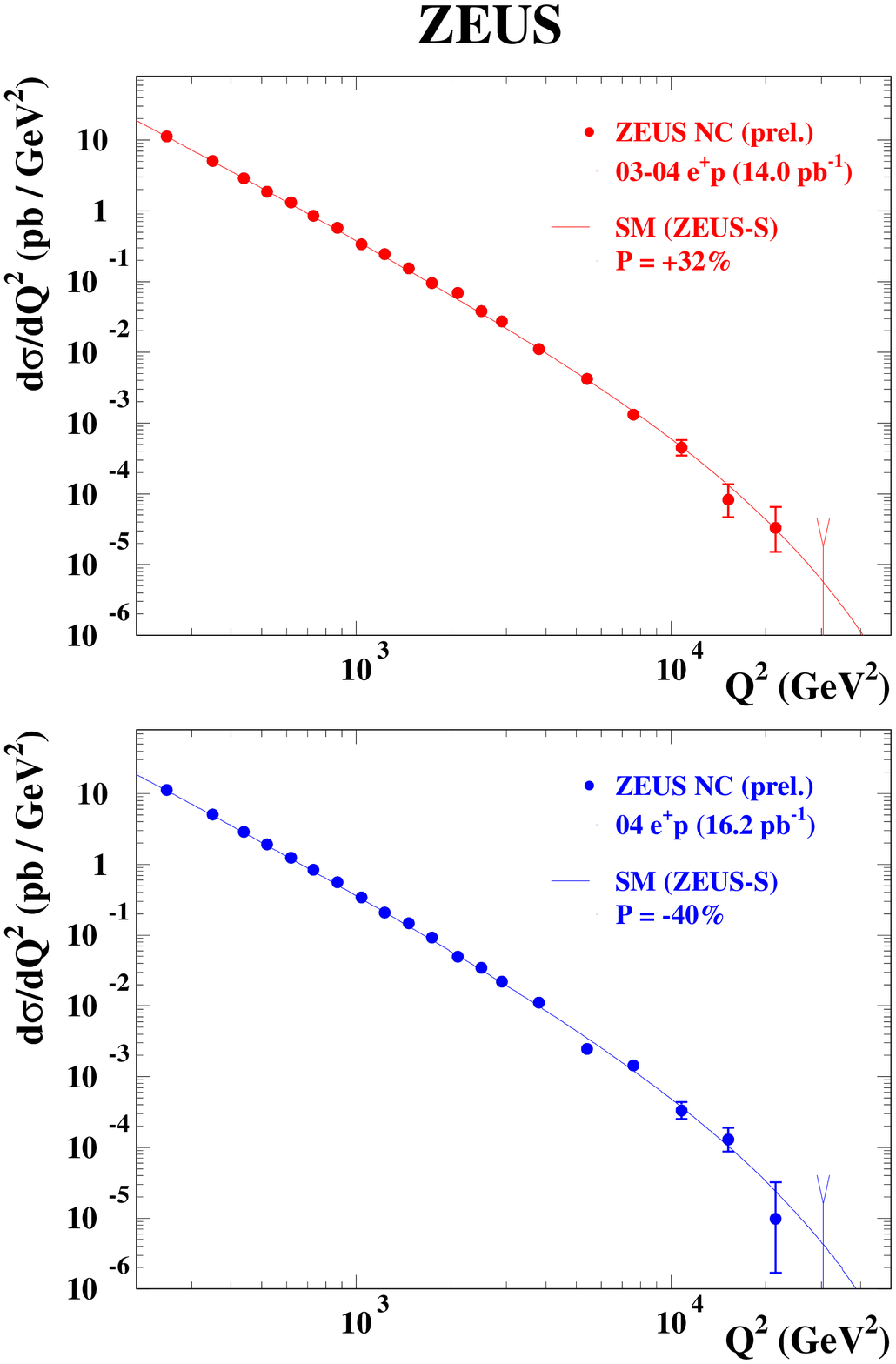,width=\textwidth}
  \end{minipage}\hfill
  \begin{minipage}[c]{0.495\textwidth}
\psfig{figure=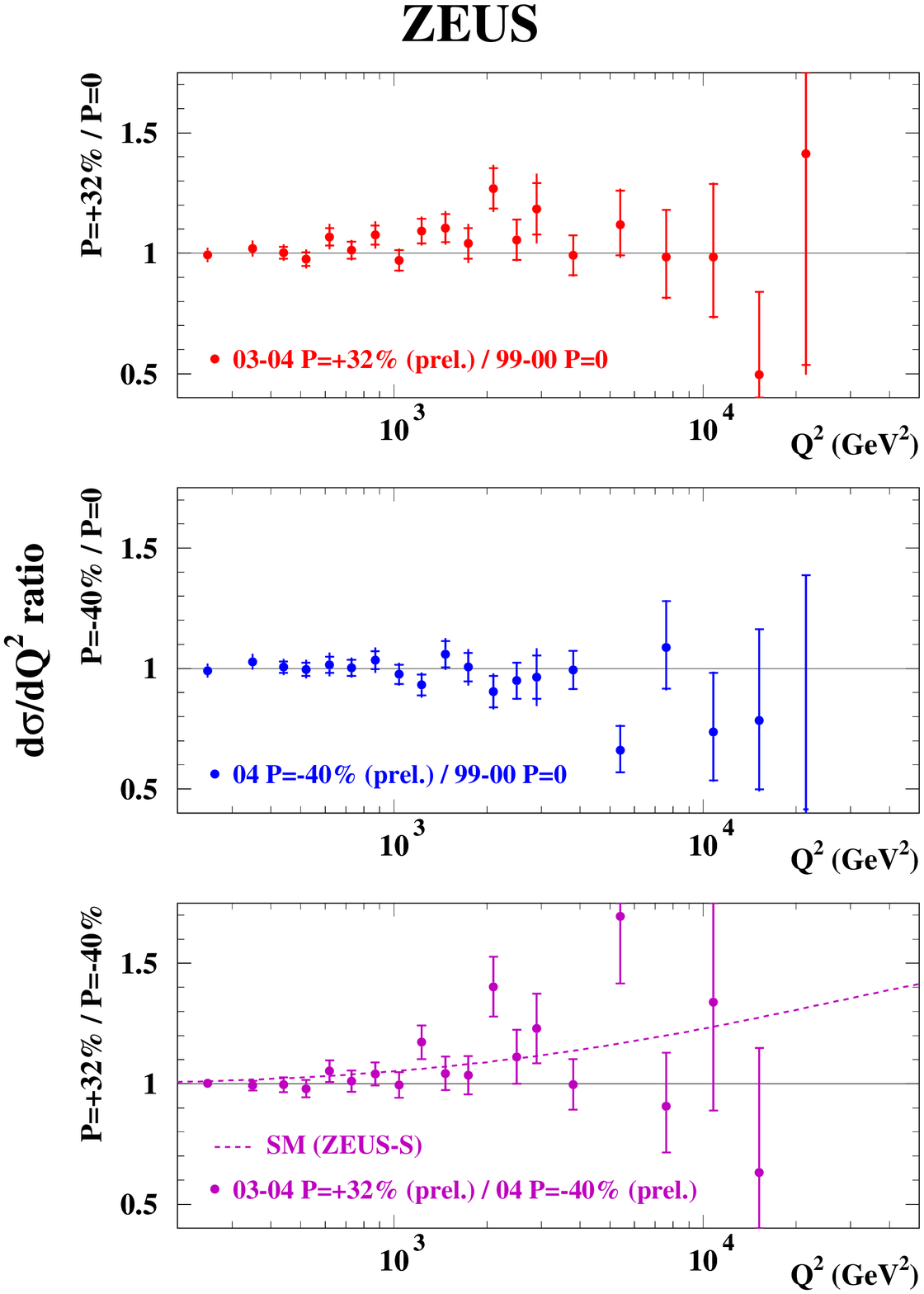,width=\textwidth}
  \end{minipage}%
\caption{The ZEUS $e^+p$ DIS NC cross-sections $d\sigma / dQ^2$ for positive (upper left) and negative
(lower left) polarised data. The curves show the SM prediction evaluated using the ZEUS-S PDFs.
On the right the ratios of the NC DIS cross sections $d\sigma / dQ^2$ 
for the positively polarised data to unpolarised data (upper), 
negatively polarised data to unpolarised data (middle) and 
positively polarised data to negatively polarised data are shown.
The dashed curve shows the SM prediction evaluated using the ZEUS-S PDFs.
\label{fig:nc-xsec}}
\end{figure}
The cross-sections $d\sigma/dQ^{2}$ for NC DIS for positive and negative 
longitudinal polarisations as measured by ZEUS are shown in Fig.~\ref{fig:nc-xsec} 
together with the ratios of the cross sections for the positive and negative 
longitudinal polarisations to the unpolarised results~\cite{Chekanov:2003vw}. 
Also shown is the ratio of the cross sections for 
the positive and negative longitudinal polarisations. 
Only statistical uncertainties were considered when taking ratios
of the positively and negatively polarised cross sections. 
In taking ratios to the unpolarised cross sections the 
systematic uncertainties were considered uncorrelated with those of the polarised cross sections. 
The measurements are well described by the SM evaluated using the ZEUS-S PDFs and consistent with
the expectations of the electroweak SM for polarised NC DIS, 
although the statistical precision of the current data set does not allow 
the polarisation effect to be conclusively observed.

\end{document}